\documentclass [12pt,a4paper,draft]{article}

\usepackage{times}

\DeclareFontFamily{OT1}{times}{}
\DeclareFontShape {OT1}{times}{m }{n }{ <-> ptmr }{}
\DeclareFontShape {OT1}{times}{bx}{n }{ <-> ptmb }{}
\DeclareFontShape {OT1}{times}{m }{it}{ <-> ptmri}{}
\DeclareFontShape {OT1}{times}{bx}{it}{ <-> ptmbi}{}
\usepackage{amsmath}
\usepackage{amsfonts}
\usepackage{amssymb}
\usepackage{latexsym}
\newcommand{\cl}{C \kern -0.1em \ell} 
\newcommand{\DEF}{:=}                 

\newcommand{\CON}{\overline}          

\newcommand{\Scal}{\mathbb{S}}        
\newcommand{\Vect}{\mathbb{V}}        

\newcommand{\BRA}{\langle\kern -.2em\langle} 
\newcommand{\KET}{\rangle\kern -.2em\rangle} 

\newcommand{\REA}{\operatorname{Re}}  
\newcommand{\IMA}{\operatorname{Im}}  

\newcommand{\Q}{[\hspace{1.mm}]} 
\newcommand{\A}{(\hspace{.5mm})} 

\newcommand{\ASS}{\thickapprox}       
\newcommand{\ADJ}{\dagger}            
\newcommand{\INV}{{-1}}               
\begin{document}

\title{{\vskip -3cm} Explicit closed-form parametrization of $SU(3)$ and $SU(4)$ in terms of complex quaternions and elementary functions\\  ~\\ {\large \emph{Submitted to J. Math. Phys.}}}

\author{André Gsponer}

\date{ISRI-02-05 ~~ \today}

\maketitle

\begin{abstract}

Remarkably simple closed-form expressions for the elements of the groups $SU(n)$, $SL(n,\mathbb{R})$, and $SL(n,\mathbb{C})$ with $n=2$, 3, and 4 are obtained using linear functions of biquaternions instead of $n \times n$ matrices.  These representations do not directly generalize to $SU(n>4)$.  However, the quaternion methods used are sufficiently general to find applications in quantum chromodynamics and other problems which necessitate complicated  $3 \times 3$ or $4 \times 4$ matrix calculations.

\end{abstract}

\section{Introduction}

The continuous and compact groups $SU(3)$ and $SU(4)$ play important roles in physics, especially in gauge theories of elementary particles interactions, and in the classification of nuclear and hadronic states and resonances.  However, while the three-parameters group $SU(2)$ is sufficiently simple to be easily formulated in various elegant forms using $2 \times 2$ matrices or quaternions \cite{GILMO1974-}, the published parametrizations of the eight-parameters group $SU(3)$ are comparatively much more complicated, e.g., references $[2-12]$. 

In this paper, simple and explicit closed-form expressions for the elements of the groups $SU(n)$, $SL(n,\mathbb{R})$, and $SL(n,\mathbb{C})$ with $n=2$, 3, and 4 are built using linear functions of biquaternions instead of $n \times n$ complex matrices.\footnote{To avoid the frequent use of the adjective \emph{complex} we will use the prefix \emph{bi-} that was suggested by Hamilton to qualify complex numbers, vectors, and quaternions, i.e., elements of  $\mathbb{B}=\mathbb{C}\otimes\mathbb{H}$.} These expressions include ``Lie-type'' representations (in which the full set of Lie generators is explicitly used to write the general element of the group) and ``Euler-angles'' representations (in which only a subset of the Lie generators is used). So far as we know, our representations are new.

Since linear biquaternion functions of biquaternions correspond to $4 \times 4$ complex matrices, the biquaternion representations presented in this paper are in fact equivalent to $4 \times 4$ matrix representations.  This means, when representing for example $SU(3)$ with biquaternions, that the extraneous fourth dimension can be used as an auxiliary component to make the calculation of the group elements easier with quaternions than with $3 \times 3$ matrices.  Thus, the essence of the method applied in this paper is equivalent to the seminal idea that lead to the discovery of quaternions by Hamilton in 1843, namely the concept that the multiplication and division of vectors (i.e., triplets of numbers) are only possible if an auxiliary number (the scalar part of the quaternion) is introduced to enable the calculation.

An interesting aspect of this method, i.e., \emph{using quaternions to provide a kind of ``algebraic continuation'' to facilitate  $3 \times 3$ matrix calculations}, is that it is general.  Not only are the calculations necessary to obtain the triplet representation of $SU(3)$ elementary (and requiring only a basic knowledge of quaternion algebra) but the same method may be used with similar efficiency to build  higher dimensional representations such as the octet, the decuplet, etc.  Moreover, the method may be applied to various calculations in quantum chromodynamics, which are known to be quite difficult, and to related problems in particle physics. Finally, as it happens with the second of the two representation of $SU(3)$ given in this paper, some quaternion expressions written for three-dimensional vectors keep their simplicity when generalized to four dimensions.  This is how a concise expression is obtained for the fifteen parameter group $SU(4)$.

Of course, the possibility of representing small groups such as $SU(2)$, $SO(3)$, and $SO(4)$ with quaternions is well known.  For instance, since the three parameters defining a $SU(2)$ group element can be assembled in the vector part of a real quaternion of unit norm $\exp(\tfrac{1}{2}\alpha\vec{a})$, the multiplicative group of such quaternions provides a representation of $SU(2)$.  In this representation a column vector of complex numbers $(x,y)$ is mapped onto a real quaternion $q \in \mathbb{H}$ by the expression
$$
   q =  \bigl(\REA(x) + \IMA(x) e_1\bigr)
     +  \bigl(\REA(y) + \IMA(y) e_1\bigr) e_2 \eqno(1)
$$
where $e_1$ and $e_2$ are any two out of the three quaternion units, and the corresponding $2 \times 2$ matrix of $SU(2)$ is represented by the real quaternion function\footnote{The round parentheses in the notation $F{\A}$ designate that $F$ is a function whose argument is conceived to occupy the place marked by $\A$, while the square brackets $\Q$ are conceived to mark the position to be occupied by a quaternion within a quaternion monomial, e.g., $AB \Q CD$. This suggestive notation due to Hamilton \cite[p.359]{HAMIL1891-} was later promoted by Conway and Synge \cite{CONWA1945A,SYNGE1972-}.}
$$
             M_{SU(2)}{\A} = e^{\tfrac{1}{2}\alpha\vec{a}}{\Q}   \eqno(2)
$$
where $\vec{a}=\sum_{k=1}^3 a_ke_k$ is a unit vectors and $\alpha \in \mathbb{R}$. Then
$$
        q' = M_{SU(2)}(q) = e^{\tfrac{1}{2}\alpha\vec{a}} q   \eqno(3)
$$
is an $SU(2)$ transformation such that the invariant quaternion form $q'\CON{q'}=q\CON{q}$ is the counterpart of the Hermitian form $(x',y')^\ADJ(x',y') = (x,y)^\ADJ(x,y) = x^*x + y^*y$ which is invariant in the standard $2 \times 2$ complex matrix formalism. Unfortunately, this particularly elegant (and explicitly real) realization of $SU(2)$ does not generalize to $SU(3)$ and $SU(4)$.

   In the case of $SO(3)$ and $SO(4)$ one has the representations
$$
M_{SO(3)}{\A}= e^{\tfrac{1}{2}\alpha\vec{a}} {\Q} e^{-\tfrac{1}{2}\alpha\vec{a}}
                                                             ~~,  \eqno(4)
$$
and
$$
M_{SO(4)}{\A} = e^{\tfrac{1}{2}\alpha\vec{a}} {\Q} e^{-\tfrac{1}{2}\beta\vec{b}}
                                                             ~~,  \eqno(5)
$$
which clearly show that these groups have three and six parameters,  respectively.

   Expressions $(2)$, $(4)$, and $(5)$ are elementary examples of linear functions of biquaternions.  As will be recalled in Sec.~2, any complex $4 \times 4$ matrix $\mathcal{M}$ can always be replaced by the linear function $M{\A}: \mathbb{B} \rightarrow \mathbb{B}$ isomorphic to $\mathcal{M}: \mathbb{C}^4 \rightarrow \mathbb{C}^4$.  In such linear functions the argument is inserted in empty spaces which may occur (because of the noncommutativity of quaternions) anywhere within a linear expression such as, e.g., $M{\A} = A{\Q}B + {\Q}C$, where $A, B,$ and $C$ are quaternions.  Therefore, to avoid possible confusion, a quaternion used as an \emph{operator} (e.g., a linear function) will always be written $Q{\Q}$ or ${\Q}Q$, while a quaternion used as an \emph{operand} (or a number in an expression) will always be written $Q$. Moreover, to reduce the proliferation of parentheses, and to make long expressions more readable, the symbol $\odot$ will be used to separate operators according to the obvious composition rule, e.g., $A{\Q}B \odot C{\Q} \odot {\Q}D = A\bigl( C\bigl( {\Q}D \bigl) \bigr)B = AC{\Q}DB$.

The representations $(4)$ and $(5)$ of the orthogonal groups $SO(3)$ and $SO(4)$ will be of direct use in generalizing the well-known Euler-angles representation of $SU(2)$ to $SU(3)$ and $SU(4)$.  This will be possible because of {\bf Lanczos's fundamental decomposition theorem} which states that \emph{any arbitrary, nonzero and possibly rectangular, real \emph{(complex)} matrix $C$ can be written as the product of an orthogonal \emph{(unitary)} matrix $A$, a positive diagonal \emph{(phase)} matrix $D$, and the transpose \emph{(adjoint)} of a second orthogonal \emph{(unitary)} matrix $B$,} i.e.,  \cite{LANCZ1958-}
$$
                  C = A D B^t    ~~.                             \eqno(6)
$$
Therefore, any unitary matrix may be written as
$$
                  U = O_1 D_\phi O_2^t    ~~                     \eqno(7)
$$
where $O_1, O_2$ are two general orthogonal matrices such as $(4)$ or $(5)$, and $D_\phi$ a diagonal phase matrix.  However, while this decomposition is essentially unique because the two orthogonal matrices combined with the diagonal matrix have exactly the right number of independent parameters required to represent $SU(n)$, it is not the only possible one.  For instance, we will begin with a quaternionic parametrization of $SU(3)$ corresponding to the decomposition
$$
                  U = O_A D_\phi U_S    ~~                       \eqno(8)
$$
where $O_A$ corresponds to an orthogonal matrix and $U_S$ to a particular unitary matrix.
 
The outline of the paper is as follows. In Sec.~2 the linear quaternion functions equivalent to general symmetric, antisymmetric, and diagonal $3 \times 3$ and $4 \times 4$ matrices are given.  In Sec.~3 the exponential maps of these functions are calculated.  In Sec.~4 these maps are used in a Lie-type representation of $SU(3)$ corresponding to the decomposition $(8)$.  In Sec.~5 Euler-angles representations of $SU(2)$, $SU(3)$, and $SU(4)$ are built according to the decomposition $(7)$.  Finally, in Sec.~6, the obtained representations are compared to various non-quaternion representations, some hints for building further quaternions representations are given, and some advantageous features of quaternion representations are highlighted.\footnote{
Throughout this paper a number of equivalent representations of the same groups will be written. In order to simplify the notation, all the parameters, whether scalars \{$\alpha, \beta, \gamma, ...$\} or vectors \{$\vec{a}, \vec{b}, \vec{c}, ...$\}, will be represented by the same symbols  even though they may correspond to different numerical values.}

\section{Linear functions and Conway operators}

The most common language for expressing linear functions is that of matrices with real or complex number elements.  However, if hypercomplex numbers such as quaternions are used, it is possible to express any linear function corresponding to one $4 \times 4$ matrix (sixteen complex numbers) by a unique linear combination of sixteen  elementary quaternion operators.  Since the quaternion algebra is non-commutative, these so-called ``Conway operators'' are of the type $e_n{\Q}e_m$ where the empty space corresponds to the position of the argument, and where $e_k$ ($k=1,2,3$) are the three quaternion units and $e_0=1$ is the ordinary scalar unit \cite{CONWA1945A,SYNGE1972-}.

   For instance, if the biquaternion $Q=\sum_{n=0}^3 x_n e_n=x_0+\vec{x}$ with $x_n \in \mathbb{C}$ is used to represent a  $4 \times 1$ column vector, the quaternion form of the general $\mathbb{C}^4 \rightarrow \mathbb{C}^4$ linear function $ M{\A}: Q \mapsto Q'=\sum_{n=0}^3 x{_n}' e_n$ is then
$$
    Q' = M(Q) = \sum_{n=0}^{3} \sum_{m=0}^{3}  z_{nm}e_n[Q]e_m   \eqno(9)
$$
where $z_{nm} \in \mathbb{C}$, and $Q, Q' \in \mathbb{B}$. While this expression may seem cumbersome at first, its power stems from the fact is that many particular linear functions which are important in mathematics or physics have remarkably simple and elegant forms when they are expressed in terms of Conway operators.

For example, the linear function 
$$
 A_{\{a_1,a_2,a_3\}}{\A} = \tfrac{1}{2} \Bigl( \vec{a}{\Q} - {\Q}\vec{a} \Bigr) ~~,
                                                                \eqno(10)
$$
corresponds to the upper-left $3 \times 3$ antisymmetric matrix
$$
\begin{pmatrix}
 0    & -a_3 & +a_2 & 0 \\
 +a_3 &   0  & -a_1 & 0 \\
 -a_2 & +a_1 &  0   & 0 \\
 0    &   0  &  0   & 0 \\
\end{pmatrix}                    ~~. \eqno(10')
$$
The linear function $(10)$ directly generalizes to $\tfrac{1}{2} \bigl( \vec{a}{\Q} - {\Q}\vec{b} \bigr)$ which,  written as
$$
 A_{\{\vec{B},\vec{E}\}}{\A} =
   \tfrac{1}{2} \Bigl( \bigl(\vec{E}+i\vec{B}\bigr){\Q} + {\Q}\bigl(\vec{E}-i\vec{B}\bigr) \Bigr)
                                                          ~~,   \eqno(11)
$$
corresponds to the general $4 \times 4$ antisymmetric matrix
$$
\begin{pmatrix}
 ~0   & -iB_3 & +iB_2 & E_1 \\
 +iB_3 &  ~0  & -iB_1 & E_1 \\
 -iB_2 & +iB_1 & ~0   & E_1 \\
 -E_1 & -E_2 & -E_3 & 0   \\
\end{pmatrix}                     \eqno(11')
$$
where we have intentionally taken the combinations $\vec{E}\pm i\vec{B}$ for the parameters to stress that this matrix has the same form as the electromagnetic field tensor.

Similarly, the linear function
$$
 D_{\{d_1,d_2,d_3\}}{\A}  = 
\tfrac{1}{2}\Bigl(d_1 e_1{\Q}e_1 + d_2 e_2{\Q}e_2 + d_3 e_3{\Q}e_3 \Bigr) \eqno(12)
$$
corresponds to the general $4 \times 4$  traceless diagonal matrix
$$  \frac{1}{2}
\begin{pmatrix}
 -d_1+d_2+d_3 &      0      &      0      &        0       \\
       0      & d_1-d_2+d_3 &      0      &        0       \\
       0      &      0      & d_1+d_2-d_3 &        0       \\
       0      &      0      &      0      & -(d_1+d_2+d_3) \\
\end{pmatrix}                                                      \eqno(12')
$$
which shows that $(12)$ corresponds to an upper-left $3 \times 3$ traceless diagonal matrix when the three numbers $d_k$ are subject to the condition $d_1+d_2+d_3=0$.

Finally, the general diagonal-less $3 \times 3$  symmetric matrix can also be neatly expressed in quaternions.  Starting from the linear function $\tfrac{1}{2}\lambda\vec{s}{\Q}\vec{s}$, where $\vec{s}$ is a unit vector, one has simply to subtract a $4 \times 4$ diagonal function in order to remove the diagonal terms. Therefore,
$$
 \lambda S_{\{s_1,s_2,s_3\}}{\A} = \lambda\Bigl( \tfrac{1}{2}\vec{s}{\Q}\vec{s}
                           -  D_{\{s_1^2,s_2^2,s_3^2\}}{\A}   \Bigr)  \eqno(13)
$$
corresponds to the diagonal-less symmetric matrix
$$ \lambda
\begin{pmatrix}
   0    & s_1s_2 & s_1s_3 & 0 \\
 s_1s_2 &   0    & s_2s_3 & 0 \\
 s_1s_3 & s_2s_3 &   0    & 0 \\
   0    &   0    &   0    & 0 \\
\end{pmatrix}
= \lambda
\begin{pmatrix}
  0  & t_3 & t_2 & 0 \\
 t_3 &  0  & t_1 & 0 \\
 t_2 & t_1 &  0  & 0 \\
  0  &  0  &  0  & 0 \\ 
\end{pmatrix}                        ~~.     \eqno(13')
$$
Thus, the components of the vector $\vec{s}$ can be calculated from the matrix elements $t_k$ by the formulas $s_1^2=t_2 t_3/t_1$, $s_2^2=t_3 t_1/t_2$, $s_3^2=t_1 t_2/t_3$.

   Unfortunately, to obtain the quaternion equivalent of the general diagonal-less $4 \times 4$ symmetric matrix, it is not enough to replace one of the vectors $\vec{s}$ in $(13)$ by a different unit vector $\vec{u}$ because the resulting term $\tfrac{1}{2}\lambda\vec{s}{\Q}\vec{u}$ would have only five independent parameters, just like the electromagnetic energy-moment tensor $\tfrac{1}{2}\bigl(\vec{E}+i\vec{B}\bigr){\Q}\bigl(\vec{E}-i\vec{B}\bigr)$.  The correct generalization of $(13)$ is
$$
 S_{\{\sigma,\nu,\vec{s},\vec{u}\}}{\A} = 
  \mathbb{D}_{\{\sigma,\nu\}}{\A} \odot
          \Bigl( \tfrac{1}{2} \vec{s}{\Q}\vec{u} 
                -  D_{\{s_1u_1,s_2u_2,s_3u_3\}}{\A} \Bigr) \odot
  \mathbb{D}_{\{\sigma,\nu\}}{\A} \eqno(14)
$$
where the diagonal function $\mathbb{D}_{\{\sigma,\nu\}}{\A} = \sigma \Scal{\Q} + \nu \Vect{\Q}$ produces a rescaling between the scalar and vector parts of the operand.  As a result, equation $(14)$ is somewhat cumbersome and not very useful.

In conclusion --- since any matrix can be expressed by the sum of a symmetric, an antisymmetric, and a diagonal matrix --- the general $3 \times 3$ traceless matrix can, according to $(10 - 13)$, always be represented by the expression
$$
 M{\A} = \lambda S_{\{s_1,s_2,s_3\}}{\A}
              + A_{\{a_1,a_2,a_3\}}{\A}
              + D_{\{d_1,d_2,d_3\}}{\A} ~~.\eqno(15)
$$
Moreover, if the diagonal term in the symmetric function $(13)$ is merged with the diagonal function $(12)$, the final expression
$$
   M{\A} =         \tfrac{1}{2} \lambda  \vec{s}  {\Q}  \vec{s}
                 + \tfrac{1}{2} \Bigl( \vec{a}{\Q}  -  {\Q}\vec{a} \Bigr)
   + D_{\{d_1 -\lambda  s_1^2,
          d_2 -\lambda  s_2^2,
          d_3 -\lambda  s_3^2\}}{\A}     \eqno(16)
$$
provides a neat quaternion representation of a traceless $\mathbb{C}^3 \rightarrow \mathbb{C}^3$ linear function $\vec{v} \mapsto \vec{v'}$ characterized by the parameters $\lambda$, $s_k$, $a_k$, and $d_k$ subject to the conditions $s_1^2+s_2^2+s_3^2=1$ and $d_1+d_2+d_3=0$.  However, this expression does not conveniently generalize to arbitrary traceless $\mathbb{C}^4 \rightarrow \mathbb{C}^4$ linear functions --- an illustration of the loss of power of the quaternion method  when going from a three to a four dimensional problem.

\section{Exponential maps of $S{\A}$, $A{\A}$, and $D{\A}$}

In order for the functions $S{\A}$, $A{\A}$, and $D{\A}$ to be useful representations of the generators of the Lie algebras corresponding to $SU(3)$ and $SU(4)$ it is necessary that these quaternion functions lead to elementary analytical expressions when their respective exponential maps are summed to go from infinitesimal to finite group transformations. Moreover, since  $SU(3)$ and $SU(4)$ are compact groups, it is necessary that these expressions are themselves ``compact,'' i.e., expressible in terms of trigonometric functions only.

For instance, in the case of the antisymmetric function $A{\A}$, we have to calculate
$$
        \operatorname{EXP} \alpha A_{\{ a_1, a_2,  a_3\}}{\A} = \operatorname{EXP} \tfrac{1}{2}\alpha\Bigl(\vec{a}{\Q}-{\Q}\vec{a}\Bigr) \eqno(17)
$$
where $\vec{a}$ is a unit vector, $\alpha$ a real parameter, and (following the convention of Gilmore) the symbol $\operatorname{EXP}$ designates the Taylor series corresponding to the Lie expansion of the group near the origin \cite{GILMO1974-}.  Obviously, the result is well-known. Indeed, because the commutator $\bigl[\vec{a}{\Q} , {\Q}\vec{a}\bigr]=0$, and $\bigl(\vec{a}{\Q}\bigr)^2= \bigl({\Q}\vec{a}\bigr)^2=-1{\Q}$, we get   
$$
\operatorname{EXP} \tfrac{1}{2} \alpha \Bigl( \vec{a}{\Q} - {\Q}\vec{a} \Bigr) = \exp(\tfrac{\alpha}{2}\vec{a}){\Q}\exp( -\tfrac{\alpha}{2}\vec{a})     \eqno(18)
$$
which is nothing but equation $(4)$, the celebrated Olinde-Rodrigues formula for spatial rotations in quaternion form \cite{CAYLE1848-}.

The case of the diagonal function $D{\A}$ is also trivial. In effect, since the Conway operators in the diagonal function $(12)$ commute with each other, it comes
$$
  \operatorname{EXP} i D_{\{\delta_1,\delta_2,\delta_3\}}{\A}  = 
$$
$$
    \exp\Bigl(\delta_1 \tfrac{i}{2} e_1{\Q}e_1 \Bigr) \odot
    \exp\Bigl(\delta_2 \tfrac{i}{2} e_2{\Q}e_2 \Bigr) \odot
    \exp\Bigl(\delta_3 \tfrac{i}{2} e_3{\Q}e_3 \Bigr)               \eqno(19)
$$ 
where, because $\bigl(e_k{\Q}e_k\bigr)^2 = +1{\Q}$ instead of $-1{\Q}$, $d_k$ has been replaced by $i\delta_k$ in order to obtain a result that is ``compact'' when $\delta$ is real. 

The case of the diagonal-free symmetric function $S{\A}$ is slightly more complicated. This is because the exponential map
$$
 \operatorname{EXP} i\beta S_{\{b_1, b_2, b_3\}}{\A} = 
 \operatorname{EXP} i\beta\Bigl( \tfrac{1}{2} \vec{b}{\Q}\vec{b}
                           -D_{\{b_1^2,b_2^2,b_3^2\}}{\A} \Bigr)  \eqno(20)
$$
does \emph{not} lead to a simple expression. However, the first term in the exponent does immediately lead to 
$$
\operatorname{EXP} \tfrac{1}{2} i \beta \vec{b}{\Q}\vec{b}
            = \exp \tfrac{1}{2} i \beta \vec{b}{\Q}\vec{b}         \eqno(21)
$$
where, because $\bigl(\vec{b}{\Q}\vec{b}\bigr)^2 = +1{\Q}$ when $\vec{b}$ is a unit vector, the imaginary unit $i$ has been introduced to obtain a ``compact'' result when $\beta$ is real .

Thus, while only the last two terms of $(15)$ lead to a simple exponential map, all three terms of $(16)$ have such a property. This suggests that the later representation should be used when calculating the $\operatorname{EXP}$onential of a traceless linear function.

\section{Quaternionic ``Lie-type'' representations of\\
           $SU(3)$, $SL(3,\mathbb{R})$, and $SL(3,\mathbb{C})$}

It is evident from their matrix representations $(10'-13')$ that the linear functions $A{\A}$, $S{\A}$, and  $D{\A}$ introduced in the preceding section to obtain compact expressions for the exponential maps can be related to the Lie generators of $SU(3)$ by simple algebraic expressions.  These relations are given in Table~1 for the Gell-Mann parametrization of $SU(3)$.

Therefore, according to the general theorem of Lie relating the generators of a Lie algebra to the elements of its corresponding group we have the canonical map
$$
  G_{C}{\A} = \operatorname{EXP} \Bigl(
   \alpha A_{\{     a_1,      a_2,      a_3     \}}{\A} +
   i\beta S_{\{     b_1,      b_2,      b_3     \}}{\A} +
        i D_{\{\delta_1, \delta_2, \delta_3     \}}{\A}    \Bigr) \eqno(22)
$$ 
where the exponent corresponds to the general expression $(15)$ for a trace-less linear function. Unfortunately, just like in $3 \times 3$ matrix representations, this map does not lead to a simple closed-form expression for the group elements.  But, if we use the results of the preceding section, we immediately see that if equation $(16)$ is used instead of $(15)$ we get a representation that is fully expressible in terms of elementary functions.  In effect, if using the Baker-Campbell-Hausdorf theorem \cite{GILMO1974-, WILCO1967-} the expression
$$
   \operatorname{EXP}  \Bigl(
        i\beta  \tfrac{1}{2}          \vec{b}{\Q}\vec{b} +
         \alpha \tfrac{1}{2} \bigl(\vec{a}{\Q} - {\Q}\vec{a} \bigr)   +
   i D_{\{\delta_1 - \beta b_1^2, 
          \delta_2 - \beta b_2^2,
          \delta_3 - \beta b_3^2\}}{\A}    \Bigr)                \eqno(23)
$$
is written as the composition of three EXPonential factors
$$
\operatorname{EXP} \Bigl( i\tfrac{\beta}{2}\vec{b}{\Q}\vec{b}\Bigr) \odot
\operatorname{EXP} \tfrac{\alpha}{2}\Bigl(\vec{a}{\Q}-{\Q}\vec{a}\Bigr) \odot
\operatorname{EXP} \Bigl( i  D_{\{
                   \delta_1 - \beta b_1^2, 
                   \delta_2 - \beta b_2^2,
                   \delta_3 - \beta b_3^2\}}{\A}  \Bigr)         \eqno(24)
$$
we get a non-canonical representation of $SU(3)$ which, according to $(18)$, $(19)$, and $(21)$, is simply
$$
   G_{SU(3)}{\A} = 
\exp \Bigl(   \tfrac{\beta}{2} i \vec{b}{\Q}\vec{b}              \Bigr) \odot
\exp          \tfrac{\alpha} {2} \Bigl(\vec{a}{\Q} - {\Q}\vec{a} \Bigr) \odot
\exp \Bigl(iD_{\{\delta_1,\delta_2,\delta_3,\beta,\vec{b}\}}{\A} \Bigr) \eqno(25)
$$
where
$$
\exp \Bigl(iD_{\{\delta_1,\delta_2,\delta_3,\beta,\vec{b}\}}{\A} \Bigr) =  \exp\Bigl((\delta_1 - \beta b_1^2)\tfrac{i}{2}e_1{\Q}e_1 \Bigr) \odot \hspace{3cm}
$$
$$\hspace{3cm}
 \exp\Bigl((\delta_2 - \beta b_2^2)\tfrac{i}{2}e_2{\Q}e_2 \Bigr) \odot
 \exp\Bigl((\delta_3 - \beta b_3^2)\tfrac{i}{2}e_3{\Q}e_3 \Bigr) \eqno(26)
$$
Equations $(25,26)$ are final fully explicit expressions for $SU(3)$ group elements.  All parameters are real and reduce to eight independent ones because of the conditions $a_1^2+a_2^2+a_3^2 = b_1^2+b_2^2+b_3^2 =1$ and $\delta_1+\delta_2+\delta_3=0$.  If the imaginary units $i$  are suppressed in $(25,26)$, the resulting equations give a representation of the non-compact groups $SL(3,\mathbb{R})$ when all parameters are real, and $SL(3,\mathbb{C})$ when they are complex.

For these representations it is important to remark that the order of the exponentials in $(25)$ is immaterial --- although every permutation gives another element of the  group.  In particular, the inverse element is not simply obtained by changing the signs in the exponents, the order of the exponential factors has to be reversed at the same time. For instance, 
$$
   G_{SU(3)}^\INV{\A} = 
\exp \Bigl(-iD_{\{\delta_1,\delta_2,\delta_3,\beta,\vec{b}\}}{\A} \Bigr) \odot
\exp \tfrac{\alpha}{2} \Bigl( {\Q}\vec{a} - \vec{a}{\Q}  \Bigr) \odot 
\exp  \Bigl( - \tfrac{\beta}{2} i \vec{b}{\Q}\vec{b}      \Bigr) ~~.  \eqno(27)
$$
Therefore, comparing with $(25)$, and writing ${\A}^+$ for biconjugation (i.e., the combination of imaginary and quaternion conjugations) and ${\A}^\ASS$ for function association,\footnote{\emph{Function association} (called function conjugation by  Hamilton \cite[p.555]{HAMIL1891-}), which reverses the order of all operations in a linear function, e.g., $\bigl( ab{\Q}cd \odot ef{\Q}gh   \bigr)^\ASS = gh{\Q}ef \odot cd{\Q}ab$, is defined by the scalar equation $\Scal\bigl[ F(X) ~ Y \bigr] = \Scal\bigl[X ~ F^{\ASS}(Y) \bigr]$ where $F^\ASS\A$ is the \emph{associate} of the linear function $F\A$.} the inverse of any $SU(3)$ group element can be written
$$
G_{SU(3)}^\INV{\A} =G_{SU(3)}^{+\ASS}{\A} =G_{SU(3)}^\ADJ{\A} \eqno(28)
$$
which implies that ${\A}^{+\ASS}$ is the quaternion equivalent of Hermitian conjugation~${\A}^\ADJ$.  Hence, using the definition of function association, we immediately verify that the representation $(25)$ conserves as expected the Hermitian form $\Scal[X^+Y]$, i.e., that $\Scal\bigl[ G_{SU(3)}^+(X^+) ~ G_{SU(3)}(Y) \bigr] = \Scal[X^+Y]$.

\section{Quaternionic ``Euler-angles'' representations of \\$SU(3)$ and $SU(4)$}

The most common parametrizations of finite three-dimensional rotations and spin~$\tfrac{1}{2}$ transformations are based on the particular matrix representation
$$
 \mathcal{U}_{\{\alpha,\beta,\gamma\}} = \exp(i\alpha J_3)
                                         \exp(i\beta  J_2)
                                         \exp(i\gamma J_3)   \eqno(29)
$$  
in which only two out of the three Lie generators of $SU(2)$ appear, and where $\{\alpha, \beta,\gamma\}$ are the standard Euler-angles defined in many textbooks, e.g., \cite{WIGNE1931-}.  The advantage of this representation is that the diagonal Pauli matrix $J_3$ is used twice so that the resulting matrix is of maximum simplicity, i.e.,
$$
\mathcal{U}_{\{\alpha, \beta,\gamma\}} = \pm
\begin{pmatrix}
  e^{\tfrac{1}{2}i(-\alpha-\gamma)}\cos\tfrac{1}{2}\beta &
 -e^{\tfrac{1}{2}i(-\alpha+\gamma)}\sin\tfrac{1}{2}\beta \\
  e^{\tfrac{1}{2}i(+\alpha-\gamma)}\sin\tfrac{1}{2}\beta &
 +e^{\tfrac{1}{2}i(+\alpha+\gamma)}\cos\tfrac{1}{2}\beta \\
\end{pmatrix}                                                ~~. \eqno(29')
$$
  However, if the objective is to work with a representation that can be generalized to higher dimensional unitary groups it is better to use 
$$
 \mathcal{U}_{\{\alpha, \beta,\gamma\}}' = \exp(i\alpha J_2)
                                           \exp(i\beta  J_3)
                                           \exp(i\gamma J_2)  ~~. \eqno(30) 
$$
If the first and the second quaternion vector coordinates are chosen to represent the $SU(2)$ doublet as $q=x e_1+y e_2$, and if the corresponding Pauli matrices are expressed  by means of formulas $(10,10')$ and $(12,12')$, the quaternion equivalent of $(30)$ is
$$
 U_{\{\alpha, \beta,\alpha\}}{\A} =
e^{\tfrac{1}{2}\alpha{e_3}}{\Q}e^{-\tfrac{1}{2}\alpha{e_3}} \odot                          \exp i\tfrac{\beta}{2} \bigl({e_1}{\Q}{e_1}
                           - {e_2}{\Q}{e_2}        \bigr) \odot
e^{\tfrac{1}{2}\gamma{e_3}}{\Q}e^{-\tfrac{1}{2}\gamma{e_3}}
                                                                  \eqno(30')
$$
whose simplicity is comparable to that of the matrix $(29')$. But $(30')$ has the additional advantage to correspond to Lanczos's decomposition $(7)$ with two orthogonal function of the form $(4)$, and a unitary phase diagonal function of the form $(12)$, i.e., $D_{\phi} = \exp \bigl( iD_{\{\beta, -\beta, 0\}}{\A}\bigr)$.  Hence, according to $(7)$, the generalization of $(30')$ to $SU(3)$ is therefore
$$
 U_{SU(3)}{\A} =
e^{\tfrac{1}{2}\alpha\vec{a}}{\Q}e^{-\tfrac{1}{2}\alpha\vec{a}}  \odot                             \exp \Bigl( iD_{\{\beta, \gamma, -\beta-\gamma\}}{\A} \Bigr) \odot
e^{\tfrac{1}{2}\delta\vec{b}}{\Q}e^{-\tfrac{1}{2}\delta\vec{b}}
                                                                     \eqno(31)
$$
which has eight parameters: four phase $\alpha, \beta, \gamma, \delta$, and four angles in the two unit vectors $\vec{a}$ and $\vec{b}$.

   The next level of generalization is also trivial: $SU(4)$ is obtained by replacing the two $SO(3)$ factors by two $SO(4)$ functions of the form $(5)$, and by using for $D_\phi$ the general trace-less diagonal function $(12)$.  It comes
$$
 U_{SU(4)}{\A} =
e^{\tfrac{1}{2}\alpha\vec{a}}{\Q}e^{-\tfrac{1}{2}\beta\vec{b}}  \odot                             \exp \Bigl( iD_{\{\gamma, \delta, \epsilon\}}{\A} \Bigr) \odot
e^{\tfrac{1}{2}\psi\vec{c}}{\Q}e^{-\tfrac{1}{2}\eta\vec{d}}
                                                                     \eqno(32)
$$
which has fifteen parameters: seven phases and eight angles.

Finally, expression $(30')$ can be rewritten in a general form by replacing the quaternion units by three orthogonal unit vectors $\vec{u}$, $\vec{v}$, and $\vec{w}=\vec{u}\times\vec{v}$.  For the $SU(2)$ doublet  $q=x\vec{u}+y\vec{v}$, we have then the formula
$$
 U_{SU(2)}{\A} =
e^{\tfrac{1}{2}\alpha\vec{w}}{\Q}e^{-\tfrac{1}{2}\alpha\vec{w}} \odot                            \exp i\tfrac{\beta}{2} \bigl(\vec{u}{\Q}\vec{u}
                           - \vec{v}{\Q}\vec{v}        \bigr)  \odot
e^{\tfrac{1}{2}\gamma\vec{w}}{\Q}e^{-\tfrac{1}{2}\gamma\vec{w}}
                                                              ~~.   \eqno(33)
$$

In summary, expressions $(33)$, $(31)$, and $(32)$ are fully general and coordinate-free representations of the groups $SU(n)$ for $n=2$, 3, and 4.  Moreover, if the imaginary factor $i$ is suppressed we get similar representations for the corresponding groups $SL(n,\mathbb{R})$ or $SL(n,\mathbb{C})$ when all the parameters are either real or complex. Finally, it is easy to generalize or specialize between these formulas, as well as to isolate various unitary, orthogonal, or Abelian subgroups.

\section{Discussion}

Considering that the groups $SU(3)$ and $SU(4)$ depend on eigth or fifteen  parameters, it is remarquable that explicit expressions such as $(25,26)$ and $(31,32)$ are possible, and that only trigonometric functions and a few quaternion multiplications are needed to calculate any group element.  However, due to the fact that infinitely many equivalent parametrizations are possible for any Lie group, the merits of our quaternionic parametrizations have to be judged in view of possible applications, and in comparison with other published parametrizations, that we will briefly review.

\noindent {\bf ``Lie-type'' representations}

A first class of representations comprise those which use explicitly the full set $\{\lambda_n\}$ of the Lie generators so that any group element is expressed by a map $G{\A} = G\bigr(\pi_1\lambda_1{\A}, ..., \pi_N\lambda_N{\A}\bigl)$ where $\pi_n$ is the parameter associated with the infinitesimal generator $\lambda_n{\A}$, and $N$ the total number of generators, i.e., $N=8$ for $SU(3)$.

\emph{Canonical form:} The group element is expressed as the EXPonential of the linear combination ~ $\pmb{\pi} \cdot \pmb{\lambda}{\A}$, i.e.,
$$
  G_C{\A} = 
  \operatorname{EXP}\bigl(\pmb{\pi}\cdot\pmb{\lambda}{\A}\bigr) \DEF
  \operatorname{EXP}\bigl(\sum_{k=1}^N \pi_n\lambda_n{\A}\bigr)  ~~.\eqno(34)
$$
Unfortunately, the infinite series of terms implied by the EXP symbol is usually very difficult to be summed in closed form.  In particular, while the series lead to expression $(2)$ in the case of $SU(2)$, a comparably simple formula has not yet been obtained for $SU(3).$ 

\emph{Non-canonical forms:} As a consequence of the Baker-Campbell-Hausdorff theorem \cite{GILMO1974-, WILCO1967-} it is possible to break-down the canonical form into a product of EXPonentials such as
$$
 G_N{\A} = \operatorname{EXP} \bigl(\sum_{n=1}^{m_1}\pi_n\lambda_n{\A}\bigr)
                              \odot ... \odot 
           \operatorname{EXP} \bigl(\sum_{n=m_k}^{N}\pi_n\lambda_n{\A}\bigr)
                                                                     \eqno(35)
$$
with the hope that the EXPonentials could be summed in a closed form. This is what we have done in equation $(24)$ in order to get a product of \emph{three} EXPonentials (the three diagonal exponentials are counted as a single one since they are trivial). In fact, the best that has been achieved so far in the case of $SU(3)$ was to find a closed form expression for a \emph{two} EXPonentials representation in which the non-diagonal generators were merged in one EXPonential that was laboriously evaluated using a combination of matrix recurrence relations and Laplace transform techniques \cite{RAGHU1989-}.  However, it is not impossible (by using more involved techniques then those of the present paper) that a similar result could be achieved with quaternions --- and that even the canonical form could be summed this way.

\emph{Product form:} An extreme non-canonical form is obtained by taking advantage of the normalization $\bigl(\lambda_n{\A}\bigr)^2 = \pm1{\A}$ to factorize $(34)$ into a product of $N$ exponentials, i.e.,
$$
 G_P{\A} = 
       \bigodot_{n=1}^{N} \operatorname{EXP} \bigl(\pi_n\lambda_n{\A}\bigr)
     = \bigodot_{n=1}^{N} \exp \bigl(\pi_n\lambda_n{\A} \bigr) ~~.\eqno(36)
$$
However, this ``brute-force'' method has the disadvantage that the corresponding finite transformations's parameters $\pi_n$ have generally no clear physical or geometrical interpretation.  This is why the product form is mostly used in application where the intrinsic meaning of the parameters is not essential, as was the case in the first published parametrizations \cite{MURNA1962-, CHACO1966-, NELSO1967-, HOLLA1969-}. 

\emph{Basis elements (or exponential) form:} An alternative to the canonical form, which is possible for matrices as well as for quaternions, is to expand the elements of the group into a sum over the basis elements $1{\A}$ and $\lambda_n{\A}$,
$$
G_B{\A} = u_0(\pmb{\pi}){\A} + i\sum_{n=1}^{N} u_n(\pmb{\pi})\lambda_n{\A}
                                                                 ~~. \eqno(37)
$$
For $SU(2)$ the coefficients are simply $u_0=\cos(\tfrac{1}{2}\alpha)$ and $u_n=\sin(\tfrac{1}{2}\alpha)a_n$~.  But for $SU(3)$ the coefficients can only be calculated after a cubic equation has been solved for every group element \cite{MACFA1968-}.

\emph{Hamilton-Cayley form:} In applications such as the quark model or strong interactions, where the so-called $SU(3)$ ``octet-vector'' $\pmb{\pi}$ has a direct interpretation \cite{MICHE1973-}, it is useful to expand the elements of the group into a power series in $\pmb{\pi} \cdot \pmb{\lambda}{\A}$~. Because of the Hamilton-Cayley theorem this series has three terms for $SU(3)$,  
$$
     G_H{\A} = C_1(\pmb{\pi}){\A}
         + C_2(\pmb{\pi})\bigl(\pmb{\pi} \cdot \pmb{\lambda}{\A}\bigr)
         + C_3(\pmb{\pi})\bigl(\pmb{\pi} \cdot \pmb{\lambda}{\A}\bigr)^2
                                                                  ~~, \eqno(38)
$$
so that only three coefficients have to be calculated instead of nine as in the exponential form. These coefficients were found to be rather complicated at first \cite{ROSEN1971-}, but it was later established that they could be expressed in terms of $SU(3)$ invariants \cite{MACFA1980-}.  More recently, somewhat simpler formulas were obtained by expressing the  $SU(3)$ group elements in terms of two orthonormal vectors instead of just the octet-vector \cite{BINCE1990-, KUSNE1995-}, and further progress is likely \cite{WEIGE1997-}.  Interestingly, if the octet-vector is expressed in terms of the quaternion linear functions that appear in the exponent of the canonical representation $(23)$, 
$$
  \pmb{\pi} \cdot \pmb{\lambda}{\A} =
        i\beta  \tfrac{1}{2}          \vec{b}{\Q}\vec{b} +
         \alpha \tfrac{1}{2} \bigl(\vec{a}{\Q} - {\Q}\vec{a} \bigr)   +
         iD_{\{\delta_1,\delta_2,\delta_3,\beta,\vec{b}\}}{\A} 
                                                                 ~~, \eqno(39)
$$
it is possible to get a quaternion equivalent of $(38)$.  However, the physical interpretation will not be the same as with the Gell-Mann octet-vector because the parameters (i.e., $\vec{a}, \vec{b}, \alpha, \beta,$ and $\delta_n$; see Table 1) will not relate to the quark model, or to the gluon creation/destruction operators of quantum chromodynamics, in the usual way.  But this may precisely be the main interest of the quaternion approach: an opportunity for new interpretations.

\noindent {\bf ``Euler-angles'' representations}

In Euler-angles representations only a subset $\{\lambda_{l(n)}\} \subset \{\lambda_{n}\}$ of the Lie generators is used to formulate the general element of the group:  
$$
G_E{\A} =
      \operatorname{EXP} \bigl(\sum_{n=1}^{m_1}\pi_n\lambda_{l(n)}{\A}\bigr)
                              \odot ... \odot 
      \operatorname{EXP} \bigl(\sum_{n=m_k}^{N}\pi_n\lambda_{l(n)}{\A}\bigr)
                                                               ~~.  \eqno(40)
$$
For instance, in the case of $SU(n)$, the minimal number of generators needed to generated the whole set is $2(n-1)$, i.e., two for $SU(2)$ as in $(29-30')$. Since that number is four for $SU(3)$, many early parametrizations used just four generators \cite{NELSO1967-, HOLLA1969-}.  In this respect, our own Euler-angles representation $(31)$ is not minimal since it uses five different generators.  But this number could easily be reduced to four if the two $SO(3)$ factors were replaced by their Euler-angles counterparts.  However, doing so would lead to a more complicated representation that would spoil the main advantage of $(31)$, namely to show that in line with the usual interpretation of Euler-angles in ordinary three-space rotations, a general finite $SU(3)$ ``rotation'' in complex three-space is obtained by making first a three dimensional rotation around an axis $\vec{b}$, then a diagonal rotation, and finally a second three dimensional rotation around an axis $\vec{a}$.

\noindent {\bf Specific advantages of quaternion representations}

As we have just seen with the Euler-angles formulation of $SU(3)$, quaternion representations have a number of didactical advantages.  For example, the use of quaternions forces a ``natural'' grouping of parameters into scalars and vectors which may have a geometrical or a physical interpretation, and which may lead to various formal or algebraic simplifications.

In this perspective, the grouping of the symmetrical, antisymmetrical, and diagonal matrix elements into separate linear quaternion function may lead to further geometrical insight into the geometry of $SU(3)$ transformations.  For instance, with regards to the representation $(25)$, the arguments of the second and third exponentials correspond to the so-called ``q-'' and ``r-'' octet-vectors of Michel and Radicati \cite{MICHE1973-}. What is then the geometric meaning of the argument of the first exponential?

Concerning the applications of the quaternion representations given in this paper, possibly the most obvious ones are in the field of particle and nuclear physics, as already mentioned in the introduction.  However, further applications in mathematics and physics may stem from the general need for spherical harmonics and their generalization to the ``complex rotations'' corresponding to $SU(3)$ and $SU(4)$, e.g., \cite{NELSO1967-}.  In this respect, since many practical calculations with ordinary spherical harmonics are linked to Euler-angles representations of the rotation group, e.g., \cite{WIGNE1931-}, the  representations $(31 - 33)$ can be of great help to generalize known $SU(2)$ results to $SU(3)$ and $SU(4)$.

Finally, compared to matrices, a general advantage of quaternion representations or formulations is that they are coordinate-free.  For instance, in matrix representations, finding subgroups consists of isolating ``blocks'' or particular combinations of lines and rows, and then possibly of making a change of basis to get the general case.  On the other hand, in quaternion representations such as $(31-33)$, it is possible to ``continuously'' specialize/generalize between groups and subgroups, and thus to obtain, if not all, at least a large fraction of all related groups and subgroups, directly in their most general form.

\section{Acknowledgments}
It is a pleasure to thank Dr.\ Jean-Pierre Hurni for stimulating discussions and mathematical guidance through the intricacies of the theory of Lie groups.

\begin{table}
\begin{center}
\begin{tabular}{|l|l|l|l|} 		
\hline
\multicolumn{2}{|c|}{\raisebox{+0.2em}{{\bf Matrices ~ $\pmb{\Longleftrightarrow}$ ~ Quaternions ~~ \rule{0mm}{5mm}}}} \\ 
\hline
$\pi_1=-2\beta b_1 b_2$   & $a_1=+\pi_7/2\alpha$ \rule{0mm}{5mm} \\
$\pi_2=+2\alpha_3$        & $a_2=-\pi_5/2\alpha$ \rule{0mm}{5mm} \\
$\pi_3=\delta_1-\delta_2$ & $a_3=+\pi_2/2\alpha$ \rule{0mm}{5mm} \\
$\pi_4=-2\beta b_3 b_1$   & $b_1=\sqrt{\pi_1\pi_4/2\beta\pi_6}$ \rule{0mm}{5mm} \\
$\pi_5=-2\alpha a_2$      & $b_2=\sqrt{\pi_1\pi_6/2\beta\pi_4}$ \rule{0mm}{5mm} \\
$\pi_6=-2\beta b_2b_3$    & $b_3=\sqrt{\pi_4\pi_6/2\beta\pi_1}$ \rule{0mm}{5mm} \\
$\pi_7=+2\alpha a_1$      & $\delta_1= -\tfrac{1}{2}(\pi_8/\sqrt{3}+\pi_3)$ \rule{0mm}{5mm} \\
$\pi_8=\sqrt{3}(\delta_1+\delta_2)$ & $\delta_2= -\tfrac{1}{2}(\pi_8/\sqrt{3}-\pi_3)$ \rule{0mm}{5mm} \\
\hline
\multicolumn{2}{|c|}{\raisebox{+0.2em}
{{ $\delta_1+\delta_2+\delta_3=0$ \rule{0mm}{5mm}}}} \\ 
\hline
\multicolumn{2}{|c|}{\raisebox{+0.2em}
{{ $\alpha=\tfrac{1}{2}\sqrt{\pi_2^2+\pi_5^2+\pi_7^2}$ \rule{0mm}{5mm}}}} \\ 
\hline
\multicolumn{2}{|c|}{\raisebox{+0.2em}
{{ $\beta=\tfrac{1}{2}\sqrt{|\pi_1\pi_4/\pi_6|+|\pi_1\pi_1/\pi_4|+|\pi_4\pi_6/\pi_1|}$ \rule{0mm}{5mm}}}} \\ 
\hline
\end{tabular}
\end{center}
\caption{\emph{Relation between the parameters of the quaternion $SU(3)$ representation (22) and the parameters $\pi_n$ of the standard representation  based on the Gell-Mann matrices $\lambda_n$. }}    \label{`tab:0.1'}
\end{table}

\end{document}